\documentclass[acmsmall,screen,nonacm]{acmart}

\AtBeginDocument{%
  }

\setcopyright{acmlicensed}
\copyrightyear{2018}
\acmYear{2018}
\acmDOI{XXXXXXX.XXXXXXX}
\acmConference[Conference acronym 'XX]{Make sure to enter the correct
  conference title from your rights confirmation email}{June 03--05,
  2018}{Woodstock, NY}
\acmISBN{978-1-4503-XXXX-X/2018/06}

\usepackage{graphicx}

\usepackage[utf8]{inputenc}
\usepackage{listings}
\usepackage{xcolor}
\usepackage[most]{tcolorbox} 
\usepackage{enumitem}
\usepackage{subfigure}
\usepackage{svg}

\usepackage{booktabs}
\usepackage{multirow}
\usepackage[table,xcdraw]{xcolor}

\usepackage{adjustbox}

\definecolor{cbblue}{rgb}{0.2,0.4,0.9}
\definecolor{cbgreen}{rgb}{0,0.6,0.3}
\definecolor{cbred}{rgb}{0.9,0.2,0.2}
\definecolor{cbgray}{gray}{0.4}
\definecolor{bgcolor}{rgb}{0.98,0.98,0.98}

\tcbset{highlight style/.style={
    colframe=#1!80,
    colback=#1!10,
    colupper=black,
    boxrule=0.8pt,
    left=2pt,
    right=2pt,
    top=1pt,
    bottom=1pt,
    arc=3pt,
    boxsep=0pt
}}

\lstdefinestyle{mypython}{
    language=Python,
    basicstyle=\ttfamily\small,
    numbers=left,
    numberstyle=\tiny\color{gray},
    numbersep=3pt,
    backgroundcolor=\color{bgcolor},
    keywordstyle=\color{blue}\bfseries,
    showstringspaces=false,
    escapeinside={(*@}{@*)},
    breaklines=true
}

\begin{document}


\title{Code Clone Detection via an AlphaFold-Inspired Framework}


\author{Changguo Jia}
\email{jiachangguo@stu.pku.edu.cn}
\affiliation{%
  \institution{School of Computer Science, Peking University}
  \institution{Key Laboratory of High Confidence Software Technologies, Ministry of Education}
  \city{Beijing}
  \state{}
  \country{China}
}

\author{Yi Zhan}
\email{zhanyi@stu.pku.edu.cn}
\affiliation{%
  \institution{School of Computer Science, Peking University}
  \institution{Key Laboratory of High Confidence Software Technologies, Ministry of Education}
  \city{Beijing}
  \state{}
  \country{China}
}

\author{Tianqi Zhao}
\email{zhaotq@zgclab.edu.cn}
\affiliation{%
  \institution{Zhongguancun Laboratory}
  \city{Beijing}
  \state{}
  \country{China}
}

\author{Hengzhi Ye}
\email{hzye@stu.pku.edu.cn}
\affiliation{%
  \institution{School of Computer Science, Peking University}
  \institution{Key Laboratory of High Confidence Software Technologies, Ministry of Education}
  \city{Beijing}
  \state{}
  \country{China}
}

\author{Minghui Zhou}
\authornote{Minghui Zhou is the corresponding author (zhmh@pku.edu.cn).}
\email{zhmh@pku.edu.cn}
\affiliation{%
  \institution{School of Computer Science, Peking University}
  \institution{Key Laboratory of High Confidence Software Technologies, Ministry of Education}
  \city{Beijing}
  \state{}
  \country{China}
}



\begin{abstract}
Code clone detection, which aims to identify functionally equivalent code fragments, plays a critical role in software maintenance and vulnerability analysis. 
Substantial methods have been proposed to detect code clones. However, they struggle to extract high-level program semantics directly from a single linear token sequence, leading to unsatisfactory detection performance. 
A similar single-sequence challenge has been successfully addressed in protein structure prediction by AlphaFold through learning rich representations from single-sequence inputs. 
Motivated by the successful resolution of the shared single-sequence challenge by AlphaFold, as well as the sequential similarities between proteins and code, we leverage AlphaFold for code clone detection. 
In particular, we propose AlphaCC, which represents code fragments as token sequences and adapts AlphaFold's sequence-to-structure modeling capability to infer code semantics. 
The pipeline of AlphaCC goes through three steps. 
First, AlphaCC transforms each input code fragment into a token sequence and, motivated by AlphaFold's use of multiple sequence alignment (MSA), novelly uses a retrieval-augmentation strategy to construct an MSA from lexically similar token sequences. 
Second, AlphaCC adopts a modified attention-based encoder based on AlphaFold to model dependencies within and across token sequences. 
Finally, unlike AlphaFold’s protein structure prediction task, AlphaCC computes similarity scores between token sequences through a late interaction strategy and performs binary classification to determine code clone pairs. 
Comprehensive evaluations on three datasets, particularly two semantic clone detection datasets, show that AlphaCC consistently outperforms all baselines, demonstrating strong semantic understanding.
AlphaCC further achieves strong performance on instances where tool-dependent methods fail, highlighting its tool-independence. 
Moreover, AlphaCC maintains competitive efficiency, enabling practical usage in large-scale clone detection tasks.
\end{abstract}

\maketitle

\section{Introduction}

Code clones, which refer to code fragments that are identical or equivalent in functionality, are frequently introduced through copy-and-paste practices or repeated implementation of similar program logic~\cite{baker1995finding, kim2005empirical} in software engineering. 
While such reuse can reduce engineering effort and accelerate software development, it may also introduce hidden maintenance risks. 
Such risks materialize when duplicated code originates from buggy source code, as undetected defects propagate across the codebase, leading to accumulating technical debt and increasing maintenance cost~\cite{kim2017vuddy,shan2023gitor,xu2024dsfm}. 
Furthermore, many buggy clones are not exact copies but modified versions of the original code, making the underlying bugs harder to detect and eliminate~\cite{li2012cbcd,ebrahimi2019hmm}. 
For example, the OpenSSL Heartbleed vulnerability (CVE-2014-0160)~\cite{cve_detail} affected a wide range of systems due to direct reuse or partial cloning of the OpenSSL library~\cite{walden2020impact}. 
To this end, \textbf{code clone detection} is essential for identifying and addressing issues that may exist in semantically equivalent code fragments distributed throughout the codebase. 

Various approaches have been proposed for code clone detection. From the perspective of code representation, existing methods can be divided into two main categories: token-based and intermediate-representation (IR)-based~\cite{hua2020fcca, feng2024machine}. 
IR-based methods transform source code into structured intermediate representations, which typically take the form of Abstract Syntax Trees (ASTs) or program graphs. 
Existing approaches~\cite{jiang2007deckard, zhang2019novel, xu2024dsfm, yu2023graph, zhao2018deepsim, hua2020fcca, wu2020scdetector} leverage such representations to effectively detect \emph{semantic clones}~\cite{wang2023comparison}, namely code fragments that perform the same or very similar functionality but differ syntactically~\cite{yu2019neural}. 
However, IR-based methods rely on third-party analyzers for structural extraction and suffer from low efficiency. 
In practice, limitations in the maturity and maintenance of third-party analyzers---such as insufficient support for certain programming language versions---can prevent IR-based methods from operating (see Section~\ref{sec:robust}). 
Considering legacy software using older language versions, IR-based methods exhibit limited generalization and robustness. 
In contrast, token-based methods operate directly on the lexical representation of source code, where programs are treated as token sequences~\cite{baker1997parameterized, ankali2021development, feng2024machine}. 
Without complex structural extraction pipelines, token-based methods are generally more efficient and tool-independent; however, they sacrifice the ability to capture code semantics~\cite{sajnani2016sourcerercc, kamiya2002ccfinder}. 
In summary, existing clone detection methods suffer from either limited capability for semantic clone detection or strong reliance on third-party analyzers. 
These limitations stem from a fundamental challenge in code clone detection: extracting high-level program semantics \textbf{directly} from a single linear token sequence. 

A closely related challenge has long been studied in protein structure prediction: extracting three-dimensional structural information directly from a single linear protein sequence. 
In both domains, the \textbf{shared single-sequence challenge} arises: learning high-level representations from a single linear sequence. 
Recent advances in protein structure prediction---most notably AlphaFold~\cite{jumper2021highly}---demonstrate that this shared challenge can be effectively overcome. By learning rich high-level representations from a single protein sequence, AlphaFold provides a successful solution to this long-standing limitation~\cite{chen2024ai, wang2024machine}. 

Beyond the shared challenge, we further observe \textbf{sequential similarities} between proteins and code from a representational perspective. Both can be represented as linear symbolic sequences---amino acid sequences and token sequences, respectively. Each sequence is composed of a limited set of symbols, for example, the 20 standard amino acids in proteins or the finite set of programming tokens in code. 

Motivated by AlphaFold's success in addressing the shared single-sequence challenge, together with the sequential similarities between proteins and code, we introduce a new framework to address the single-sequence challenge in code clone detection. The proposed framework adopts a \textbf{token-based representation} without dependence on third-party analyzers (that often get obsolete over time), and leverages the \textbf{sequence-to-structure modeling capability} of AlphaFold to capture code semantics.

For an input protein sequence, since the limited information it contains is insufficient to reliably infer higher-level structural properties, AlphaFold first constructs a \emph{multiple sequence alignment} (MSA). 
By retrieving homologous protein sequences, MSA enriches the input representation with co-evolutionary information~\cite{rao2021msa}. 
Then, in order to model complex sequence dependencies, an attention-based encoder, Evoformer, is employed, enabling information exchange both within individual sequences and across homologous sequences~\cite{hu2022exploring}. 
The subsequent decoder module in AlphaFold, which maps learned representations to three-dimensional coordinates, is specific to protein structure prediction and is therefore not considered here. 
The protein modeling process in AlphaFold inspires the design of a code clone detection framework that combines MSA construction and an attention-based encoder to enrich token representations and improve the capability to detect semantic clones.

In this paper, we propose AlphaCC, an AlphaFold-inspired framework for code clone detection. 
AlphaCC proceeds through three main stages: a token semantic enhancer, an attention-based Codeformer encoder, and a classification and optimization module.
The token semantic enhancer aims to enrich token representation of code fragments. 
First, inspired by the MSA construction of AlphaFold, for each input token sequence, several lexically similar token sequences are retrieved from large-scale codebase together with their code context to construct a \emph{multiple sequence alignment of code} (Code MSA). 
To the best of our knowledge, this is \textbf{the first time} a retrieval-augmented strategy is applied to code clone detection. 
Then, AlphaCC introduces a novel mechanism that projects each token category into a type-specific semantic space, enabling the extraction of fine-grained, type-aware semantic features. 
Finally, an additional attention mechanism is applied to integrate information across these semantic spaces, enabling interactions between different token types. 
Building on these enriched token embeddings, an attention-based encoder adapted from AlphaFold's Evoformer, termed Codeformer, is designed to integrate inner-sequence attention and cross-sequence attention, thereby capturing dependencies both within individual token sequences and across multiple token sequences. 
At the final stage, the classification and optimization module computes fine-grained similarity scores between code fragments via late interaction and performs binary classification to determine whether a given pair constitutes a clone pair. 
To supervise this process, a margin-based loss function is employed as the optimization objective, increasing scores for semantically similar code fragment pairs while decreasing scores for dissimilar ones. 
It should be emphasized that the final stage is specific to the code clone detection task and is irrelevant to AlphaFold.

To evaluate the effectiveness, tool-independence, and efficiency of AlphaCC, we conduct experiments on three widely used code clone detection benchmarks: GCJ~\cite{gcj}, BigCloneBench~\cite{bigclonebench}, and OJClone~\cite{mou2016convolutional}, among which GCJ and OJClone focus on semantic clone. 
Experimental results show that AlphaCC consistently \textbf{outperforms all baseline methods} across the three benchmarks, achieving clone detection accuracies of $97.4\%$, $96.0\%$, and $98.6\%$ on GCJ, BigCloneBench, and OJClone, respectively. 
Notably, AlphaCC is, to the best of our knowledge, \textbf{the first token-based approach} reported to surpass IR-based methods, showing strong semantic understanding.
Moreover, AlphaCC maintains strong performance on instances where tool-dependent methods fail to operate, highlighting its tool-independence. It also exhibits superior efficiency, achieving faster detection than all IR-based approaches under identical computational settings.

Overall, our contributions are as follows:
\begin{itemize}[leftmargin=*]
    \item We observe similarities between protein sequences and token sequences, and creatively adapt the core idea of AlphaFold, namely sequence-to-structure modeling, to the task of code clone detection.

    \item We propose AlphaCC, a token-based code clone detection framework that adopts token representation and, inspired by AlphaFold, incorporates multiple sequence alignment construction and a modified attention-based encoder.

    \item We conduct comprehensive experiments on three widely used benchmarks, demonstrating the effectiveness, tool-independence, and efficiency of AlphaCC.
\end{itemize}

The remainder of the paper is organized as follows. 
Section~\ref{bg_related} reviews related work on code clone detection. 
Section~\ref{protein2code} introduces the biological inspiration behind our method. 
Section~\ref{alphacc_sec} presents AlphaCC, an AlphaFold-inspired framework for code clone detection.  
Section~\ref{eval} details the experimental setup and results to evaluate the effectiveness, tool-independence, and efficiency of AlphaCC. 
Section~\ref{discussion} discusses the possible interpretations of AlphaCC and examines potential threats to validity. 
Section~\ref{conclusion} concludes this paper.

\section{RELATED WORK}\label{bg_related}

This section introduces related works on code clone detection. According to different intermediate code representations, existing code clone detection methods can be divided into two categories: token-based and IR-based.


Token-based methods typically transform source code into code tokens by lexical analysis first, then detect code clones through the comparison of tokens. Kamiya et al.~\cite{kamiya2002ccfinder} develop \textit{CCFinder} to match code snippets token-by-token using suffix-tree matching algorithm. Sajnani et al.~\cite{sajnani2016sourcerercc} implement \textit{SourcererCC} to detect code clones by building an optimized inverted index for code tokens. While token-based methods are highly efficient, their lack of syntactic information significantly limits their effectiveness in detecting semantic clones.

IR-based methods analyze programs through structured intermediate representations that reflect program structure and semantics. 
According to the type of intermediate representations they rely on, IR-based methods can be further categorized into AST-based and graph-based methods.

AST-based methods use the \emph{abstract syntax trees} of code fragments to detect code clones. Baxter et al.~\cite{baxter1998clone} split the original AST into subtrees and compare the hash code of subtrees to detect code clones. Jiang et al.~\cite{jiang2007deckard} design \textit{DECKARD} to compute characteristic vectors for AST subtrees and employ locality-sensitive hashing to group similar vectors, thereby identifying clones. Furthermore, some methods, such as \textit{CDLH}~\cite{wei2017supervised}, \textit{ASTNN}~\cite{zhang2019novel}, and DSFM\cite{xu2024dsfm}, use deep learning models to encode AST into vectors and then calculate the similarity between vectors to detect code clones. While AST-based approaches can identify semantic code clones, they are often hindered by efficiency issues and limited tool-independence because of reliance on third-party AST analyzers.

For graph-based methods, they first analyze different graph representations, such as \textit{control flow graph} (CFG) and \textit{program dependency graph} (PDG), for code, and then further process the representation data to detect code clones. Krinke~\cite{krinke2001identifying} finds similar subgraphs in PDG to detect code clones. Wu et al.~\cite{wu2020scdetector} design \textit{SCDetector}, which uses centrality analysis to detect code clones from CFG. Similar to AST-based methods, graph-based approaches are effective in capturing semantic clones. 
However, graph-based approaches face low detection efficiency, and their dependence on third-party graph construction tools further restricts widespread adoption.

In summary, existing clone detection methods are constrained by insufficient ability of semantic clone detection or strong dependence on third-party analyzers. These limitations motivate the design of our framework, AlphaCC.

\begin{figure}[b]
  \centering
  \includegraphics[width=6cm]{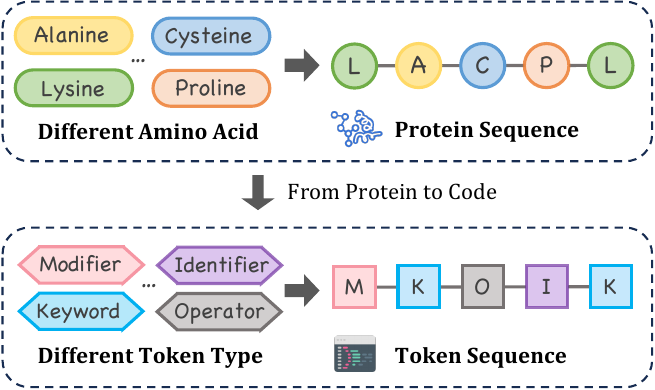}
  \caption{From Protein to Code}
  \label{protein_code}
\end{figure}

\begin{figure*}[h]
  \centering
  \includegraphics[width=\textwidth]{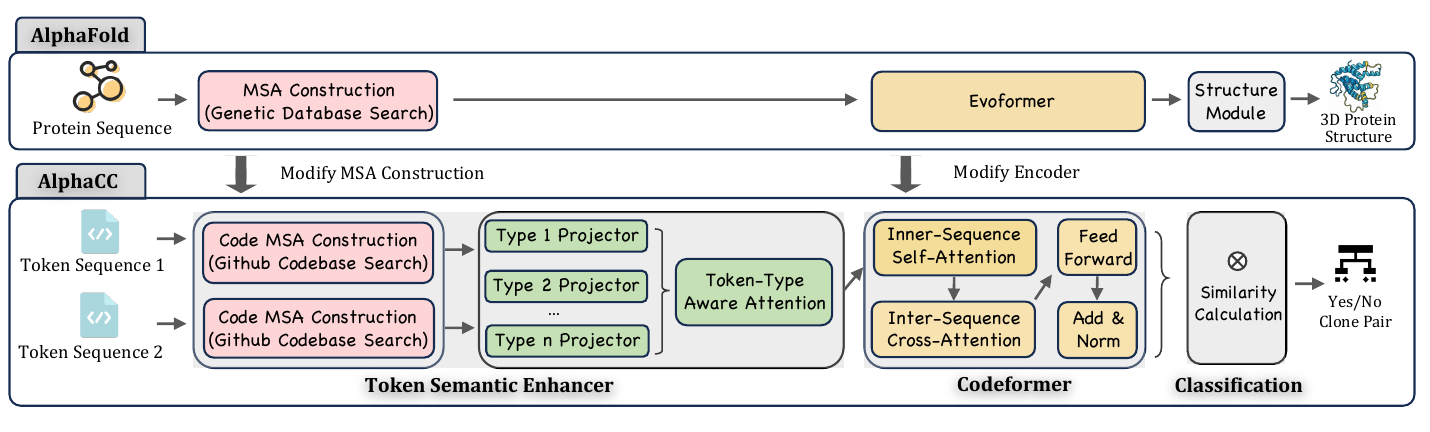}
  \caption{From AlphaFold to AlphaCC}
  \label{alphacc}
\end{figure*}

\section{Biological Inspiration: From AlphaFold to AlphaCC}\label{protein2code}

This section presents the shared design rationale underlying AlphaFold and AlphaCC, and draws an analogy between the frameworks of AlphaFold and AlphaCC to emphasize their connection and distinction. 

AlphaFold is a representative end-to-end framework that takes a protein sequence as input and predicts its three-dimensional structure. 
As illustrated in Figure~\ref{alphacc}, the overall framework consists of three stages: MSA construction, the Evoformer module, and the structure module. 
Since the structure module in AlphaFold serves as a decoder that maps learned features to three-dimensional coordinates---a stage irrelevant to code clone detection---AlphaCC focuses on adapting the first two stages to extract semantic features from token sequences. 

Accordingly, this section concentrates on the first two stages of these two frameworks. Section~\ref{subsec:ano_of_msa} discusses the adaptation of MSA construction, while Section~\ref{subsec:ano_of_encoder} describes the adaptation of Evoformer Encoder.

\subsection{MSA Construction Adaptation}\label{subsec:ano_of_msa}

The adaptation of MSA Construction is motivated by two similarities between protein structure prediction and code clone detection. 

First, there exists a sequential similarity between protein sequences and token sequences. As shown in Figure~\ref{protein_code}, both sequences share a common structural pattern: a \textbf{linear symbolic structure}. 
That is, both protein sequences and token sequences are linear sequences consisting of finite sets of discrete symbols---amino acids in the former and code tokens in the latter. 
Moreover, each symbol belongs to a predefined category. 
Protein sequences consist of $20$ standard amino acid types, such as Alanine, Cysteine, and Proline. 
Similarly, for a given programming language, code tokens fall into a finite number of categories, such as identifiers, keywords, and operators. 
Each token belongs to exactly one category, as determined by the language's lexical analyzer. 
Second, both protein structure prediction and code clone detection face an inherent challenge: a single sequence contains only limited information, which is insufficient to infer deeper properties---three-dimensional structure for proteins~\cite{shih2012evolutionary} and program functionality for code. 

To address this limitation, AlphaFold introduces MSA to aggregate homologous protein sequences into an aligned representation, enabling the capture of evolutionary and structural patterns that cannot be inferred from a single sequence alone~\cite{lupo2022protein,rao2021msa,zhang2024historical}. 
Motivated by these similarities and the role of MSA in AlphaFold, AlphaCC introduces Code MSA to enrich the semantic representation of individual token sequences. 
For each target sequence, AlphaCC retrieves functionally similar code instances from a large-scale codebase and aligns them to form a Code MSA, providing a broader context than an isolated code fragment for subsequent representation learning. 
To further enrich token representation, AlphaCC additionally introduces type-specific projections and an interaction mechanism. 
Overall, Code MSA serves as the foundation for enhancing semantic understanding in AlphaCC.

\subsection{Encoder Adaptation}\label{subsec:ano_of_encoder}

The adaptation of Evoformer encoder stems from a shared phenomenon observed in protein sequences and token sequences: both sequences exhibit \textbf{element interaction}, where the meaning of each element is context-dependent and determined by its neighboring elements. 

In protein sequences, the role of an amino acid depends on its neighboring residues. 
For instance, in the lysozyme protein sequence, the segment from positions $25$ to $35$ forms an $\alpha$-helix structure and consists of the following amino acids:
\begin{center}
    \verb|L25-G26-L27-W28-V29-A30-W31-R32-N33-R34-F35|,
\end{center}
where every letter denotes an amino acid and the accompanying number specifies its position. 
The relative positioning of amino acids such as \texttt{W28} and \texttt{W31} contributes to the formation of a hydrophobic core that stabilizes the $\alpha$-helix structure, while the local arrangement of \texttt{W28–V29–A30} creates a hydrophobic patch that is essential for local stability. 

Similarly, in a token sequence, the semantic role of each token is determined by its surrounding token context. For example, considering the C‑style loop 
\begin{center}
\verb|for(int a = 0; a < N; a++)|,
\end{center}
its corresponding token sequence is
\begin{align*}
\texttt{["for", "(", "int", "a", "=", "0", ";",}\\
\texttt{"a", "<", "N", ";", "a", "++", ")"]}.
\end{align*}
In the token sequence, token \texttt{"a"} appears multiple times, and its interpretation---whether as a declaration, a condition variable, or an increment target---relies on the surrounding tokens such as \texttt{"int"}, \texttt{"<"}, and \texttt{"++"}.

Based on this property, AlphaFold designs Evoformer to capture two crucial types of information: positional relationships of amino acids within each sequence and interactions among aligned amino acids across the entire MSA~\cite{lupo2022protein}. Inspired by Evoformer, AlphaCC introduces Codeformer, an encoder designed to model interactions among tokens both within individual token sequences and across aligned sequences in the Code MSA. Codeformer adopts a dual-attention architecture analogous to Evoformer: inner-sequence self-attention captures structural dependencies within individual token sequences, such as control and data flows, while inter-sequence cross-attention identifies co-occurrence patterns across aligned code tokens.

\section{AlphaCC}\label{alphacc_sec}

Building on cross-domain adaptations from AlphaFold, the framework of AlphaCC enables the detection of semantically equivalent code pairs across syntactic and structural variations. To support this, our method is implemented through a three-stage process, which is described in the following sections.

\subsection{Token Semantic Enhancer}

Raw token sequences lack the structural information necessary for deep semantic understanding. To address this, we design the Token Semantic Enhancer, which enriches token sequences into semantically enhanced representations through \emph{multiple sequence alignment of code} (Code MSA), enabling the model to better capture the underlying intent and functionality of the code. 

\noindent\textbf{Codebase Preparation.} 
To construct Code MSAs for token sequences, a large-scale codebase is required. As our experimental datasets (see Section~\ref{eval}) involve programs written in \texttt{Java} and \texttt{C}, we build separate codebases for each language. Specifically, we collect their top 10,000 GitHub repositories, ranked by the \textit{criticality score}, which reflects a project's overall impact and importance~\cite{criticalityscore, dou2024cc2vec}. Since the target datasets are function-level, we parse all source code from these repositories into individual functions. This results in a total of 68,093,823 functions for \texttt{Java} and 129,219,483 functions for \texttt{C}.


\begin{figure}[h]
  \centering
  \includegraphics[width=0.4\textwidth]{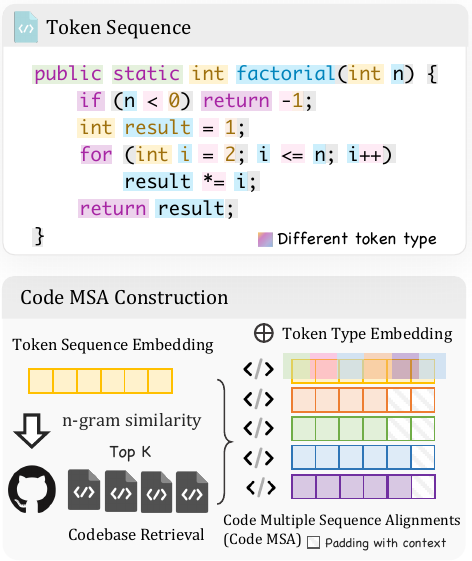}
  \caption{Construction of Code MSA}
  \label{msa_construction}
\end{figure}

\noindent\textbf{Construction of Code MSA.}
Inspired by AlphaFold’s use of \emph{multiple sequence alignment} (MSA) to reveal conserved evolutionary patterns in proteins, we introduce \emph{multiple sequence alignment of code} (Code MSA) to uncover shared structural patterns across syntactically diverse code implementations. As shown in Figure~\ref{msa_construction}, for each input code fragment, we employ an efficient $N$-gram-based similarity search to retrieve lexically similar token sequences. $N$-gram is adopted for its simplicity and efficiency in capturing recurring token subsequences, a key feature of code clones~\cite{smith2009detecting, martinez2024source, nakagawa2021nil}. It serves as a lightweight and scalable retrieval method to ensure efficient candidate alignment and broad contextual coverage. Following previous methods~\cite{nakagawa2021nil}, each sequence is represented using $5$-grams and compared via cosine similarity to retrieve the top-$k$ most similar sequences. To ensure uniformity, all sequences are standardized to a fixed length. 
All retrieved token sequences, together with the original sequence, constitute a Code MSA.

This procedure enriches the model's input by expanding it from a single implementation to a diverse set of representative patterns. 
For instance, when comparing recursive and iterative versions of a greatest common divisor (GCD) function, Code MSA provides multiple structurally similar examples from both approaches. By aligning these variations, the model learns to abstract the core computational logic, such as modulo operations and base case conditions. Simultaneously, it ignores superficial syntactic variations like variable names or control constructs. This enables a more robust understanding of functional semantics.

\begin{table}[htbp]
    \centering
    \caption{$15$ token types}
    \label{token_types} 
    \vspace{-0.5em}

    \adjustbox{max width=\textwidth}{
    \begin{tabular}{c} 
        \toprule[\heavyrulewidth] 
        \bottomrule[\heavyrulewidth] 
        \\ 
        \textquotedbl Separator\textquotedbl, \textquotedbl Identifier\textquotedbl, \textquotedbl Operator\textquotedbl, \textquotedbl Keyword\textquotedbl,
        \textquotedbl Modifier\textquotedbl, \textquotedbl DecimalInteger\textquotedbl,
        \textquotedbl BasicType\textquotedbl, \textquotedbl String\textquotedbl, \\
        \textquotedbl Boolean\textquotedbl, \textquotedbl Null\textquotedbl, \textquotedbl DecimalFloatingPoint\textquotedbl,
        \textquotedbl Annotation\textquotedbl, \textquotedbl HexInteger\textquotedbl, \textquotedbl HexFloatingPoint\textquotedbl, \textquotedbl OtherType\textquotedbl \\
        \\ 
        \toprule[\heavyrulewidth] 
        \bottomrule[\heavyrulewidth] 
    \vspace{-2em}
    \end{tabular}
    }
\end{table}

\noindent\textbf{Type-Aware Mechanism for Feature Fusion.}
To further enhance the model's structural awareness, we introduce a type-aware mechanism. As summarized in Table~\ref{token_types}, following previous research~\cite{dou2024cc2vec}, all tokens are first categorized into 14 lexical types (e.g., \textit{Identifier}, \textit{Operator}, \textit{Keyword}), along with an \textit{OtherType} that accounts for rare tokens comprising less than 1\% of the dataset. Each type is assigned a unique identifier. Based on the standardized Code MSAs, we extract the corresponding token type IDs and transform them into continuous vectors via an embedding layer. The resulting token and type embeddings are combined through element-wise addition, yielding a unified Code MSA matrix that captures both token semantic and structural information.

To model type-specific semantics, we apply a type-specific projection layer, where tokens belonging to the same lexical type are projected into a shared semantic subspace. This encourages the model to learn type-consistent representations, for example, all \textit{Operator} tokens are embedded within the same subspace, enabling the model to capture their common functional behavior. Finally, a token-type aware attention mechanism is employed to integrate information across these subspaces, effectively modeling interactions and dependencies between different token types.

\begin{figure*}[h]
  \centering
  \includegraphics[width=\textwidth, height=3.8cm]{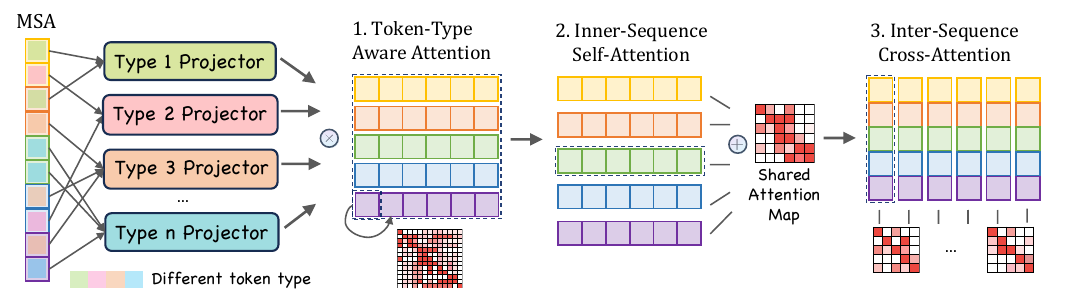}
  \vspace{-.5em}
  \caption{Pipeline of Encoding Code MSA in AlphaCC}
  \label{encoder}
  \vspace{-1em}
\end{figure*}

\subsection{Codeformer}
To interpret the enriched feature matrix, we introduce Codeformer, an encoder architecture inspired by AlphaFold’s Evoformer. As illustrated in Figure~\ref{encoder}, Codeformer employs a dual-attention mechanism that models dependencies along two aspects:

\noindent\textbf{Inner-Sequence Self-Attention.}
This component operates row-wise across the Code MSA matrix, where each row represents a single token sequence. It captures intra-procedural logic, including control flow (e.g., \textit{if-else} branches) and data flow (e.g., variable definitions and usages). Instead of computing separate attention maps for each sequence, Codeformer aggregates them into a shared attention map applied across all sequences. This design leverages the similar functional patterns often present among cloned code fragments, allowing the model to identify consistent interaction patterns that would be overlooked by isolated self-attention.

\noindent\textbf{Inter-Sequence Cross-Attention.}
This layer operates column-wise to enable interactions among tokens at aligned positions across different code fragments. It facilitates the identification of semantic equivalence across syntactic variations. For example, the model can align a \textit{for} loop in one sequence with a semantically equivalent \textit{while} loop in another, by matching their initialization, condition, and update components, even when expressed differently. This mechanism is essential for recognizing semantically similar code that has undergone significant refactoring.

Through the coordinated application of these two attention mechanisms, Codeformer generates token representations that are sensitive to both the internal logic of individual procedures and the external variation across functionally equivalent implementations.

\subsection{Classification and Optimization}

The final stage is responsible for comparing the encoded representations and optimizing the model for the code clone detection task.

\begin{figure}[h]
\vspace{-1em}
  \centering
  \includegraphics[width=7.5cm]{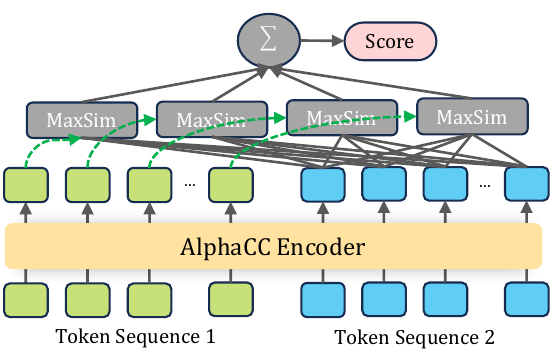}
  
  \caption{Late Interaction}
  \label{similarity}
  \vspace{-.5em}
\end{figure}

\noindent\textbf{Late Interaction.} Previous similarity computation methods typically compress entire code fragments into single vectors before comparison, a process that often results in significant information loss. To address this limitation, as shown in Figure~\ref{similarity}, AlphaCC adopts a \textit{Late Interaction} strategy~\cite{feng2020codebert, khattab2020colbert, santhanam2021colbertv2}. Instead of collapsing sequence information early, we retain the token-level representations generated by Codeformer and compute similarity scores through fine-grained, token-wise comparisons.

Specifically, we first average the token embeddings at each aligned position across all sequences in the Code MSA to obtain a single pooled representation. All token vectors are then normalized to ensure consistent scaling. Let $\mathcal{C}_1 = \{ \mathbf{t}_1^{(1)}, \mathbf{t}_2^{(1)}, \dots, \mathbf{t}_n^{(1)} \}$ and $\mathcal{C}_2 = \{ \mathbf{t}_1^{(2)}, \mathbf{t}_2^{(2)}, \dots, \mathbf{t}_n^{(2)} \}$ denote the sets of normalized token embeddings for two code fragments $\mathcal{C}_1$ and $\mathcal{C}_2$, respectively, where each $\mathbf{t}_i^{(1)}, \mathbf{t}_j^{(2)} \in \mathbb{R}^d$ and $d$ is the embedding dimension. The Euclidean distance between two token vectors is defined as:
\begin{equation}
\text{dist}(\mathbf{t}_i^{(1)}, \mathbf{t}_j^{(2)}) = \left\| \mathbf{t}_i^{(1)} - \mathbf{t}_j^{(2)} \right\|_2
\end{equation}

The overall similarity between two code fragments is computed as the mean of the minimum pairwise distances, where each token in $\mathcal{C}_1$ is matched to its nearest neighbor in $\mathcal{C}_2$:
\begin{equation}
\text{sim}(\mathcal{C}_1, \mathcal{C}_2) = \frac{1}{n} \sum_{i=1}^{n} \min_{1 \le j \le n} \text{dist}(\mathbf{t}_i^{(1)}, \mathbf{t}_j^{(2)})
\end{equation}

This late interaction approach enables fine-grained semantic alignment by directly comparing token-level representations, allowing the model to accurately identify correspondences between semantically related tokens across syntactically diverse code fragments.

\noindent\textbf{Margin Loss.}
To enable high-accuracy clone detection, we adopt \textit{Margin Loss}~\cite{you2020graph, zhan2025towards}, which encourages a more discriminative decision boundary than standard binary cross-entropy, improving generalization and reducing sensitivity to ambiguous cases.

Given the \textit{Late Interaction} similarity score $s_{ij} \in [0,1]$ between a pair of code fragments $(\mathcal{C}_i, \mathcal{C}_j)$, and a ground-truth label $y_{ij} \in \{-1, +1\}$ indicating whether the pair constitutes a clone, the loss is defined as:
\begin{equation}
\mathcal{L}_{\text{margin}} = \max \left(0, \gamma - y_{ij} \cdot (1 - s_{ij}) \right)
\end{equation}
where $\gamma > 0$ is a predefined margin hyperparameter. It enforces separation between positive and negative pairs, enhancing clone discrimination while providing a tolerance region that regularizes training and reduces overfitting.

\section{Evaluation}\label{eval}

This section presents a comprehensive evaluation of AlphaCC to rigorously assess its capabilities. Our evaluation is structured around four central research questions (RQs):
\begin{itemize}[leftmargin=*]
    \item \textbf{RQ1 (Effectiveness)\string:} How does AlphaCC's performance in code clone detection compare against state-of-the-art baseline methods?
    \item \textbf{RQ2 (Efficiency)\string:} How efficient is AlphaCC, particularly for large-scale tasks?
    \item \textbf{RQ3 (Framework Analysis):} What is the impact of AlphaCC's core components and configurations on its overall performance?
    \item \textbf{RQ4 (Robustness and Generalization):} 
    How well does AlphaCC generalize across different data distributions and cover instances unsupported by existing methods?
\end{itemize}

\subsection{Experiment Setup}

\noindent\textbf{Datasets.}
We conducted experiments on three widely used datasets. These datasets are strategically chosen to cover different programming languages and diverse clone types, ensuring a thorough and comprehensive evaluation.

\begin{itemize}[leftmargin=*]
    \item \textbf{GCJ (Google Code Jam)~\cite{gcj}:} A \texttt{Java} dataset derived from the Google Code Jam programming competition, comprising $1,664$ \texttt{Java} projects drawn from $12$ different competition problems. In line with prior work~\cite{zhao2018deepsim, xu2024dsfm}, two code snippets that deal with the same problem are treated as clones. It has been confirmed that most clone pairs in GCJ are \textbf{semantic clone}~\cite{zhao2018deepsim}. We randomly select $1,000$ code snippets for training, $332$ for validation, and $332$ for testing, resulting in nearly $600,000$ code pairs overall.

    \item \textbf{BigCloneBench~\cite{bigclonebench}:}  A massive \texttt{Java} dataset, which is constructed from IJaDataset~\cite{svajlenko2014towards} and contains over $6$ million clone pairs and $260,000$ false clone pairs. All the clone pairs are hierarchically classified into different clone types: Type-1 (T1) clones are lexically identical after comment removal and formatting normalization; Type-2 (T2) clones extend T1 by permitting systematic renaming of identifiers and replacement of literals. The remaining clone pairs are further categorized based on their similarity scores at the line and token levels: Strongly Type-3 (ST3, similarity $\in [0.7, 1.0)$) and Moderately Type-3 (MT3, similarity $\in [0.5, 0.7)$), while Type-3+4 (T4) exhibit semantic similarity with minimal textual overlap (similarity $\in [0.0, 0.5)$)~\cite{svajlenko2014towards}. T4 clones correspond to the notion of \textbf{semantic clones} as emphasized in the GCJ dataset. The scale and diversity of BigCloneBench make it well-suited for evaluating models across both syntactic and semantic clone detection challenges. Following prior work~\cite{wei2017supervised, yu2019neural, xu2024dsfm}, we randomly select $8,134$ code snippets for training, $500$ for validation, and $500$ for testing. Finally, we get $124,750$ code pairs for validation and test set and $750,500$ for training set, including $0.3\%$ of T1 clones, $0.1\%$ of T2 clones, $0.2\%$ of ST3 clones, $1.0\%$ of MT3 clones, and $98.4\%$ of T4 clones. 

    \item \textbf{OJClone~\cite{mou2016convolutional}:} A \texttt{C} language code clone dataset. OJClone is built from student submissions to $104$ programming problems~\cite{mou2016convolutional}. Prior studies~\cite{xu2024dsfm} have reported that code clone pairs from OJClone are mainly \textbf{semantic clones}, since almost all code snippets differ significantly in syntax. This makes OJClone a challenging benchmark for evaluating models’ ability to capture deep semantic similarity. Following established practice~\cite{wei2017supervised, yu2019neural, zhang2019novel}, we randomly select $6,500$ code snippets for training, $500$ for validation, and $500$ for testing, yielding $1,079,051$ training pairs and $280,875$ validation/test pairs~\cite{xu2024dsfm, yu2019neural}.

\end{itemize}

\noindent\textbf{Baselines.}
\label{baselines}
To comprehensively evaluate the effectiveness of AlphaCC, we select some state-of-the-art baselines across two types of approaches: token-based and IR-based. For IR-based methods, we mainly select from AST-based and graph-based approaches.
\begin{enumerate}[leftmargin=*]

    \item Token-based 
    \begin{itemize}[leftmargin=*]
        \item \textbf{SourcererCC}~\cite{sajnani2016sourcerercc}: A tool that identifies code clones by computing the overlap of indexed token subsequences.
        \item \textbf{NIL}~\cite{nakagawa2021nil}: A clone detection technique that employs $N$-gram token indexing to match the longest common subsequence.
        \item \textbf{Toma}~\cite{feng2024machine}: An approach that computes six similarity calculation methods based on token sequences to form feature vectors, and leverages a machine learning classifier to determine clone pairs.

    \end{itemize}
    
    \item IR (AST)-based
    \begin{itemize}[leftmargin=*]
        \item \textbf{Tamer}~\cite{hu2023fine}: A tool that splits the original AST into simple subtrees to detect code clones.
        \item \textbf{TBCCD}~\cite{yu2019neural}: An approach that utilizes tree-based convolution over token-enhanced AST to detect semantic code clones.
        \item \textbf{ASTNN}~\cite{zhang2019novel}: A model that converts AST into subtrees and uses bidirectional RNN for semantic code clone detection.
        \item \textbf{DSFM}~\cite{xu2024dsfm}: A code clone detection approach that uses GRU to capture the relationships between subtrees split from the original AST.
    \end{itemize}
    
    \item IR (Graph)-based
    \begin{itemize}[leftmargin=*]
        \item \textbf{PNIAT+PDG and PNIAT+CFG}~\cite{yu2023graph}: A technique designed to convert irregular code graph data and uses attention mechanisms to capture critical tokens within each statement in parallel to detect semantic code clones.
    \end{itemize}
    
\end{enumerate}

\noindent\textbf{Experimental Settings.}
\label{expr_set}
In our experiments, all baseline methods are configured using default settings, widely adopted hyperparameters, or values tuned on our datasets to ensure fair comparison. 
For our model, we set the number of training epochs to $1$, the learning rate to $0.0001$, the margin parameter to $0.5$, the token embedding dimension to $256$, and the top-$k$ value in Code MSA to $4$. 
All experiments are conducted on a unified computational platform equipped with a 192-core CPU, 755\,GiB of RAM, and an NVIDIA L40S GPU to ensure consistency and fairness in comparison.

\noindent\textbf{Evaluation Metrics.}
To measure the effectiveness of all the models, we adopt the widely used metrics: precision, recall, and F1 score. Precision is the proportion of correctly predicted clone pairs among all predicted clone pairs, while recall measures the proportion of correctly predicted clone pairs among all actual clone pairs. F1 score is the harmonic mean of precision and recall, providing a balanced evaluation of model performance.

\subsection{RQ1: Effectiveness Comparison}
\label{sec:effective}

To address RQ1 regarding effectiveness, we conduct a comprehensive evaluation of AlphaCC against several state-of-the-art clone detection methods across three datasets: GCJ, BigCloneBench, and OJClone, with the results shown in Table~\ref{baseline_comparison}. The best results are marked in bold, while the second-best are underlined. Some entries in Table~\ref{baseline_comparison} are marked with a dash (–) because those baseline methods could not be reproduced on OJClone. This is because OJClone is a C dataset, which some baselines do not support or were not originally designed to handle.

\begin{table}[htbp]
\centering
\caption{Comparison of different methods on three datasets}
\label{baseline_comparison}
\adjustbox{max width=\textwidth}{
\begin{tabular}{ccccccccccc}
\toprule
 & & \multicolumn{3}{c}{\textbf{GCJ}} & \multicolumn{3}{c}{\textbf{BCB}} & \multicolumn{3}{c}{\textbf{OJClone}}  \\ 
 \cmidrule(l){3-5} \cmidrule(l){6-8} \cmidrule(l){9-11}
\multirow{-2}{*}{\textbf{Type}} & \multirow{-2}{*}{\textbf{Method}} & \textbf{Precision} & \textbf{Recall} & \textbf{F1 Score} & \textbf{Precision} & \textbf{Recall} & \textbf{F1 Score} & \textbf{Precision} & \textbf{Recall} & \textbf{F1 Score} \\ 
\midrule
& SourcererCC & 74.8 & 4.2 & 7.9 & 97.5 & 0.6 & 1.2 & 27.2 & 25.2 & 26.2 \\
& NIL & 41.0 & 9.7 & 15.7 & 86.4 & 3.0 & 5.8 & 18.0 & 37.8  & 24.4              \\
\multirow{-3}{*}{Token-based} & Toma & 68.3 & 29.1            & 40.8 & 80.7 & 20.7 & 32.9 & 71.1 & 34.0 & 46.0 \\ \midrule
& PNIAT+PDG & 93.4 & 94.9 & 94.1 & 92.0 & 95.9 & 93.8 & -  & - & -   \\
\multirow{-2}{*}{IR (Graph)-based}
& PNIAT+CFG & 93.8 & 96.0 & 94.9 & 92.4 & \textbf{96.6} & 94.5 & - & - & - \\ 
\midrule
& Tamer & 60.7   & 10.3 & 17.5  & 86.0 & 2.1 & 4.1   & -  & -   & - \\
& TBCCD & 97.3   & 92.7 & 95.0  & 96.6   & 93.1 & 94.8  & 95.4   & 96.8 & 96.1  \\
& ASTNN & \textbf{97.9} & 93.0  & 95.4  & \textbf{97.9} & 92.8 & 95.3  & \underline{97.5}   & \underline{98.3} & \underline{97.9}  \\
\multirow{-4}{*}{IR (AST)-based} 
& DSFM & 97.2 & \underline{96.3} & \underline{96.8} & 95.3 & \underline{96.0}  & \underline{95.7}  & 95.7 & \textbf{99.4} & 97.5  \\ 
\midrule
\textbf{Ours} & \textbf{AlphaCC} & \underline{97.6} & \textbf{97.2} & \textbf{97.4}  & \underline{97.7}   & 94.2 & \textbf{96.0} & \textbf{99.7}   & 97.6 & \textbf{98.6}    \\    
\bottomrule
\end{tabular}
}
\end{table}

AlphaCC demonstrates superior performance across diverse benchmarks, confirming its effectiveness in code clone detection by achieving \textbf{the highest F1 scores on all three datasets}: $97.4\%$ on GCJ, $96.0\%$ on BigCloneBench, and an outstanding $98.6\%$ on OJClone. This consistent superiority across datasets with diverse programming languages and clone types highlights the effectiveness of our approach. Moreover, AlphaCC maintains \textbf{top-tier precision and recall}, demonstrating its balanced ability to both detect true clones and avoid false positives. Critically, to the best of our knowledge, AlphaCC represents the \textbf{first token-based method} reported to demonstrate superior performance over IR-based approaches in semantic clone detection.

To further explore the performance of AlphaCC, we conduct a fine-grained analysis on the challenging BigCloneBench dataset, which categorizes clone types by their syntactic similarity (Table~\ref{bcb_comparison}). AlphaCC still \textbf{demonstrates leading performance across all clone types}. 
Specifically, AlphaCC achieves a perfect F1 score of $100\%$ on T1, T2, ST3, and MT3 clones, matching the highest scores obtained by other methods. Crucially, for the most difficult T4 semantic clones, which involve significant syntactic variations, AlphaCC achieves a remarkable F1 score of 95.7\%. This result is the highest among all evaluated methods, notably surpassing strong IR-based baselines. This exceptional performance gain stems from two core architectural innovations: Code MSA, which enhances semantic representation by aligning lexically similar token sequences to reveal functional commonalities; and the Codeformer module, which jointly captures inner-sequence semantics and inter-sequence commonalities, enabling robust abstraction of underlying program logic and effectively addressing the challenge of identifying semantic clones.


\begin{table}[htbp]
\centering
\caption{Comparison of different methods in F1 Score on different clone types of BigCloneBench}
\label{bcb_comparison}
\begin{tabular}{@{}ccccccc@{}}
\toprule
\textbf{Type}                   & \textbf{Method}      & \textbf{T1}  & \textbf{T2}   & \textbf{ST3}  & \textbf{MT3}  & \textbf{T4}   \\
\midrule
\multirow{3}{*}{Token-based} & SourcererCC & 100 & 97.1 & 63.7 & 6.2  & 0    \\
                       & NIL         & 100 & 97.4 & 68.3 & 20.5 & 0    \\
                       & Toma        & 100 & 100  & 94.3 & 87.3 & 28.1 \\
\midrule
\multirow{2}{*}{IR (Graph)-based} & PNIAT+PDG   & 100 & 100  & 99.2 & 98.9 & 91.3 \\
                       & PNIAT+CFG   & 100 & 100  & 100  & 99.2 & 93.9 \\
\midrule
\multirow{4}{*}{IR (AST)-based}   & Tamer       & 100 & 100  & 98.3 & 51.2 & 2.7  \\
                       & TBCCD       & 100 & 100  & 100  & 98.3 & 94.5 \\
                       & ASTNN       & 100 & 100  & 100  & 99.7 & 95.1 \\
                       & DSFM        & 100 & 100  & 100  & 100  & 95.4 \\
\midrule
\textbf{Ours}          & \textbf{AlphaCC} & 100 & 100  & 100  & 100  & 95.7 \\
\bottomrule
\end{tabular}
\end{table}

\subsection{RQ2: Efficiency Analysis}
\label{sec:efficiency}

To investigate RQ2, we evaluate AlphaCC’s computational efficiency, a key requirement for practical deployment on large-scale clone detection tasks. We benchmark all methods on one million randomly selected code pairs from BigCloneBench, reporting the average execution time over three runs. Following prior work~\cite{yu2023graph, dou2024cc2vec}, efficiency comparisons are concentrated on three stages: data preprocessing stage, training stage (required by deep learning-based methods), and prediction stage. The evaluation results are reported in Table~\ref{efficiency_eval}.

\begin{table}[t]
\caption{Runtime performance of different methods on analyzing one million code pairs}
\label{efficiency_eval}
\adjustbox{max width=\textwidth}{
\begin{tabular}{@{}ccccc@{}}
\toprule
\textbf{Type}                         & \textbf{Method}      & \textbf{Data Preprocessing Stage} & \textbf{Training Stage} & \textbf{Prediction Stage} \\
\midrule
\multirow{3}{*}{Token-based} & SourcererCC & - & -  & 19s  \\
                             & NIL         & - & -  & 10s  \\
                             & Toma        & 75256s & 2283s & 35s \\
                             
\midrule
\multirow{2}{*}{IR (Graph)-based} & PNIAT+PDG   & 12835s & 24349s & 185s \\
                             & PNIAT+CFG   & 12643s & 24347s & 185s   \\
\midrule
\multirow{4}{*}{IR (AST)-based}   & Tamer       & - & - & 107s          \\
                             & TBCCD       & 15320s & 56195s & 5827s   \\
                             & ASTNN       & 10837s & 72417s & 8542s   \\
                             & DSFM        & 10831s & 3928s   & 475s    \\
\midrule
\textbf{Ours}         & \textbf{AlphaCC}   & 10622s & 9355s   & 76s    \\
\bottomrule
\end{tabular}
}
\end{table}

\noindent\textbf{Data Preprocessing Stage.}
AlphaCC demonstrates significantly lower data preprocessing time, completing the stage in $10,622$ seconds. 
The data preprocessing stage of AlphaCC, namely Code MSA construction, is highly parallelizable and can be executed simultaneously across multiple computing units.
Crucially, the high-dimensional vector similarity calculation allows it to be significantly accelerated with GPUs, and optimized using efficient algorithms like FAISS (Facebook AI Similarity Search)~\cite{faiss}. In contrast, baseline approaches that rely on AST or graph construction are inherently CPU-bound and predominantly serial, lacking the structural ability to leverage the concurrent processing power of GPUs, thus preventing them from matching AlphaCC's computational efficiency.

\noindent\textbf{Training Stage.}
Although methods such as DSFM exhibit shorter training durations, AlphaCC demonstrates a clear efficiency advantage over most deep learning-based approaches. AlphaCC completes training in $9,355$ seconds, significantly outperforming graph-based methods such as PNIAT+PDG ($24,349$ seconds), as well as AST-based methods like TBCCD ($56,195$ seconds) and ASTNN ($72,417$ seconds). This efficiency is attributed to AlphaCC’s lightweight design. Moreover, the training phase constitutes a one-time offline cost. When viewed in the context of AlphaCC’s fast and accurate inference capabilities, this initial training investment is therefore justified by substantial downstream benefits in performance and scalability.

\noindent\textbf{Prediction Stage.}
AlphaCC demonstrates outstanding efficiency during prediction. As shown in Table~\ref{efficiency_eval}, it is the fastest among all deep learning methods and even outperforms traditional approaches such as Tamer. 
Unlike approaches with complex model designs, AlphaCC adopts a lightweight encoder built around a few attention layers, allowing it to remain efficient while still capturing essential code semantics. 
Compared to traditional methods that are often CPU-bound and difficult to accelerate, AlphaCC is designed to fully leverage GPU hardware for efficient execution. 
This combination of architectural simplicity and GPU-friendly design enables AlphaCC to achieve efficient and scalable inference for large-scale clone detection tasks.

\subsection{RQ3: Framework Analysis}
\label{sec:Framework}

To address RQ3, we conduct a series of experiments to analyze and validate the AlphaCC's core designs. All evaluations in this section are based on the GCJ dataset for consistency.

\begin{table}[t]
\caption{Performance of AlphaCC using different code input formats}
\label{code_input_formats}
\begin{tabular}{@{}cccc@{}}
\toprule
\textbf{Method}               & \textbf{Precision} & \textbf{Recall} & \textbf{F1}   \\
\midrule
\textbf{5-sequence Code MSA (AlphaCC)}    & 97.6      & 97.2   & \textbf{97.4} \\
3-sequence Code MSA     & 97.1      & 96.4   & 96.7 \\
Original token sequence & 98.4      & 25.9   & 41.0  \\
\bottomrule
\vspace{-2em}
\end{tabular}
\end{table}

\noindent\textbf{Impact of Code MSA.}
Code MSA constitutes a central innovation in AlphaCC, aiming to enhance semantic representation by aligning lexically similar token sequences. We compare the default five-sequence Code MSA with two alternatives: a three-sequence configuration and a single token sequence. 
As shown in Table~\ref{code_input_formats}, the five-sequence MSA achieves an F1 score of $97.4\%$, outperforming both the three-sequence variant ($96.7\%$) and the single token sequence ($41.0\%$). 
This substantial improvement of Code MSA compared with the single token sequence demonstrates that exposing the model to multiple semantically related sequences encourages learning of deeper functional invariants while mitigating sensitivity to syntactic variation. This, in turn, highlights the importance and effectiveness of Code MSA in enhancing the model’s ability to capture and represent code semantics. 
In addition, the performance improvement of five-sequence Code MSA over three is not markedly significant. Therefore, constructing a three-sequence Code MSA presents a faster alternative while maintaining comparable accuracy. This efficiency makes it a viable option in situations where speed is essential, presenting a practical balance between effectiveness and efficiency.

\noindent\textbf{Ablation Study of AlphaCC.}
To validate the effectiveness of AlphaCC, we conduct an ablation study focusing on two key components: Token Semantic Enhancer and Codeformer. For the ablation of Token Semantic Enhancer, we use the original input token sequence without any enhancement. For the ablation of Codeformer, we substitute with a standard Transformer using self-attention. Results in Table~\ref{ablation_alphacc} show that the full AlphaCC model achieves the highest F1 score. 
Without Token Semantic Enhancer, F1 score of single-sequence setting significantly declined to $41.0\%$. This demonstrates that Token Semantic Enhancer indeed strengthens token-level semantics by exposing the model to multiple semantically related sequences, thereby contributing substantially to AlphaCC’s performance. 
In comparison, replacing Codeformer with a basic Transformer leads to a large decline, with F1 score reduced to $68.7\%$. This indicates that the dual-attention mechanism in Codeformer plays a crucial role in modeling deeper functional relationships within the token sequences, whereas Transformer only processes tokens at the surface level without capturing their underlying semantic correlations. Together, these results confirm that both components are essential and complementary to the overall effectiveness of AlphaCC.

\begin{table}[t]
\centering
\caption{Ablation study of AlphaCC}
\label{ablation_alphacc}
\begin{tabular}{cccc}
\toprule
\textbf{Method} & \textbf{Precision} & \textbf{Recall} & \textbf{F1} \\ \midrule
\textbf{Token Semantic Enhancer + Codeformer (AlphaCC)} & 97.6 & 97.2 & \textbf{97.4} \\ 
Token Semantic Enhancer + Transformer & 96.4 & 53.4 & 68.7 \\ 
Single token sequence + Codeformer & 98.4 & 25.9 & 41.0 \\
\bottomrule
\vspace{-2em}
\end{tabular}
\end{table}

\noindent\textbf{Ablation Study of Token Semantic Enhancer.}
To evaluate the effectiveness of the Token Semantic Enhancer, we conduct an ablation study on its two components: the token-type projection layer and the token-type aware attention mechanism. 
We consider three configurations: the full model, a variant retaining only the attention mechanism, and a baseline with both components removed. 
As shown in Table~\ref{ablation}, the full model achieves the highest F1 score. 
In comparison, removing the token-type projection layer results in a $59.6\%$ performance drop, indicating that this layer is crucial for learning discriminative token-type representations from multiple perspectives and providing richer semantic features. 
Notably, the attention-only configuration performs even worse than the variant with both components removed, with a $47.6\%$ performance drop. 
This suggests that without token-type projection, attention tends to collapse different token types into a single representation, thereby suppressing type-specific features and converging toward an averaged pattern rather than capturing fine-grained distinctions. Although subsequent modules, such as Codeformer, can model token interactions, the absence of explicit token-type information fundamentally limits semantic modeling.
Overall, these results highlight that the token-type projection layer and token-type aware attention must operate jointly to encode token-type information and facilitate semantic interactions between different types.

\begin{table}[t]
\centering
\caption{Ablation study of Token Semantic Enhancer}
\label{ablation}
\begin{tabular}{cccc}
\toprule
\textbf{Method} & \textbf{Precision} & \textbf{Recall} & \textbf{F1} \\ \midrule
\textbf{Token-Type Projection + Attention (AlphaCC)} & 97.6 & 97.2 & \textbf{97.4} \\ 
Only Attention & 55.4 & 28.7 & 37.8 \\ 
Without Projection and Attention & 85.8 & 84.9 & 85.4 \\ \bottomrule
\end{tabular}
\end{table}

\noindent\textbf{Evaluation of the Similarity and Optimization Module.}
We further examine the Classification and Optimization module by evaluating combinations of similarity measures and loss functions. Specifically, we consider three widely used similarity measures: Late Interaction, Cosine Similarity, and Euclidean Distance. In addition, we incorporate two loss functions, namely Margin Loss and Binary Cross-Entropy (BCE) Loss. 
Table~\ref{sim_and_loss} shows that the combination of Late Interaction and Margin Loss achieves the highest F1 score. Cosine Similarity performs notably worse, likely due to its insensitivity to magnitude and distributional shifts that are critical for distinguishing fine-grained semantic differences. 
It's worth noting that Euclidean Distance paired with Margin Loss also offers competitive results. Considering its lower computational overhead relative to Late Interaction, Euclidean Distance offers an efficiency–effectiveness trade-off, rendering it particularly suitable for time-critical scenarios.

\begin{table}[t]
\caption{Different Combinations of Similarity Measures and Loss Functions}
\label{sim_and_loss}
\setlength{\tabcolsep}{1mm}{
\begin{tabular}{@{}ccccc@{}}
\toprule
\textbf{Similarity Measure}         & \textbf{Loss Function} & \textbf{Precision} & \textbf{Recall} & \textbf{F1}   \\
\midrule
\textbf{Late Interaction}   & \textbf{Margin Loss}   & 97.6 & 97.2 & \textbf{97.4} \\
Cosine Similarity  & Margin Loss   & 75.1      & 97.2   & 84.7 \\
Euclidean Distance & Margin Loss   & 97.3      & 92.4   & \textit{94.8} \\
\midrule
Late Interaction   & BCE Loss      & 95.6      & 95.1   & \underline{95.3} \\
Cosine Similarity  & BCE Loss      & 66.4      & 93.1   & 77.5 \\
Euclidean Distance & BCE Loss      & 94.8      & 87.6   & 91.1 \\
\bottomrule
\end{tabular}
}
\end{table}

\subsection{RQ4: Robustness and Generalization}
\label{sec:robust}

To address RQ4, we evaluate the robustness and generalization of AlphaCC under varying data distributions and in scenarios where IR-based methods fail to operate. 

\noindent\textbf{In-domain Distribution Evaluation.}
To evaluate AlphaCC’s generalization ability, we examine its performance across different positive–negative sample ratios. This setting reflects real-world code analysis scenarios, where clone and non-clone instances often exhibit substantial class imbalance. Following common practices in imbalance evaluation, we construct datasets with multiple positive–negative ratio configurations, including 1:1, 1:5, and 1:10. As shown in Table~\ref{ratio}, AlphaCC demonstrates consistently strong performance across all imbalance settings. Even under the most extreme ratio (1:10), the model achieves an F1 score of $97.1\%$, maintaining a competitive performance. These results indicate that AlphaCC effectively captures discriminative semantic patterns regardless of class prevalence, avoiding the common bias toward majority classes. Overall, the model exhibits stable behavior under in-domain distribution shifts, further reinforcing its robustness and reliability for real-world deployment.

\begin{table}[t]
\caption{In-domain Evaluation on AlphaCC}
\label{ratio}
\begin{tabular}{@{}cccc@{}}
\toprule
\textbf{Ratio(Positive:Negative)}    & \textbf{Precision}    & \textbf{Recall}    & \textbf{F1}      \\
\midrule
 1:10 & 97.8 & 96.3 & 97.1 \\
 1:5  & 96.9 & 97.8 & 97.4 \\
 1:1  & 97.3 & 97.0 & 97.2 \\
\bottomrule
\end{tabular}
\end{table}

\noindent\textbf{Out-of-domain Distribution Evaluation.}
To assess AlphaCC’s out-of-domain generalization, we perform cross-dataset validation using three representative datasets: GCJ, BigCloneBench, and OJClone. 
Inspired by prior research practices~\cite{xu2024dsfm}, the model is trained on a source dataset and finetuned on a small subset ($10\%$) of the target dataset before testing. As reported in Table~\ref{robustness}, AlphaCC consistently maintains high performance across all transfer settings. For example, when trained on OJClone and evaluated on GCJ, the model achieves an F1 score of $96.3\%$, maintaining top-tier performance relative to all baselines. These results confirm that AlphaCC does not overfit to dataset-specific characteristics. Instead, it learns generalizable semantic representations that transfer effectively across diverse code distributions.

\begin{table}[t]
\caption{Out-of-domain Evaluation on AlphaCC}
\label{robustness}
\begin{tabular}{@{}ccccc@{}}
\toprule
\multicolumn{2}{c}{\textbf{Dataset}}   & \multicolumn{3}{c}{\textbf{Performance of AlphaCC}} \\
\cmidrule(l){1-2}\cmidrule(l){3-5}
\textbf{Training set}  & \textbf{Test set}      & \textbf{Precision}       & \textbf{Recall}       & \textbf{F1}        \\
\midrule
OJClone       & GCJ           & 98.3            & 94.3         & 96.3      \\
BigCloneBench & GCJ           & 94.3            & 92.3         & 93.3      \\
\midrule
GCJ           & OJClone       & 96.3            & 91.8         & 94        \\
BigCloneBench & OJClone       & 97.4            & 93           & 95.2      \\
\midrule
GCJ           & BigCloneBench & 88.3            & 94.2         & 91.2      \\
OJClone       & BigCloneBench & 89.6            & 91.8         & 90.7     \\
\bottomrule
\vspace{-2em}
\end{tabular}
\end{table}

\noindent\textbf{Unsupported-Instance Evaluation.}
To evaluate AlphaCC’s independence from third-party analyzers, we examine whether it can operate effectively on instances where IR-based methods fail. 
To this end, we analyze ASTs for all code snippets in the original BigCloneBench dataset, a benchmark that reflects real-world development environments. 
We use \textit{javalang}\cite{javalang}, a widely-used tool in academia, and \textit{tree-sitter}\cite{treesitter}, which has garnered great attention on GitHub, to generate ASTs.

Our analysis reveals that $16,747$ code pairs cannot be processed by \textit{javalang}, and $13,246$ pairs cannot be processed by \textit{tree-sitter}. Upon investigation, we find that \textit{javalang} supports only Java versions prior to 8 due to its lack of maintenance, while \textit{tree-sitter} supports only Java 9 and later, due to syntax conflicts with earlier language versions. Furthermore, tree-sitter produces error tokens for many instances. These cases, typically filtered out in IR-based methods evaluation, are not compatible with such approaches.

We then evaluate AlphaCC and other token-based methods on these instances. As shown in Table~\ref{token_based_pf}, all methods succeed, with AlphaCC achieving the highest F1 score of $95.8\%$. This result underscores AlphaCC's tool-independence and its strong generalization capability.

\begin{table}[t]
\caption{Performance of token-based methods on unsupported instances}
\label{token_based_pf}
\begin{tabular}{@{}cccc@{}}
\toprule
\textbf{Method}               & \textbf{Precision} & \textbf{Recall} & \textbf{F1}   \\
\midrule
SourcererCC    &  92.3   &  1.8  &   3.5  \\
NIL     &  82.6  &  2.7  &  5.2  \\
Toma    &  81.3 &  18.8  &  30.5  \\
\textbf{AlphaCC}  & 97.2  & 94.5  & 95.8 \\
\bottomrule
\vspace{-2em}
\end{tabular}
\end{table}

\section{DISCUSSION}\label{discussion}
\subsection{Interpretation and Implication}\label{interpretation}

In Section~\ref{protein2code}, we briefly outline our motivations for developing AlphaCC and explain our biological analogy-based intuition for its effectiveness. Here, we provide a more comprehensive interpretation of the factors contributing to AlphaCC's superior performance. These insights may serve as foundational principles for future research aiming to develop even more powerful code clone detection tools.

\begin{figure}[h]
    \centering
    \begin{minipage}[t]{0.45\linewidth}
        \centering
        \fbox{\includegraphics[width=\linewidth]{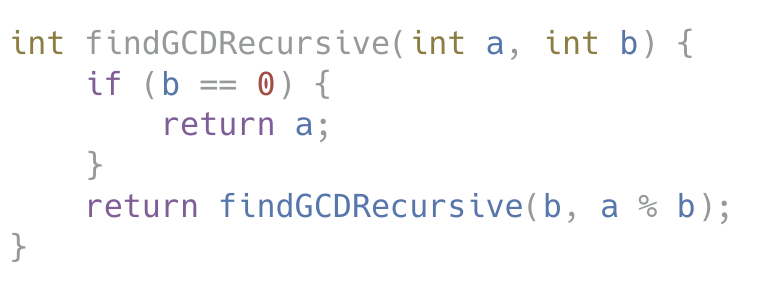}}
        \centerline{(a) Recursive Implementation}
    \end{minipage}
    \hspace{0.04\linewidth}
    \begin{minipage}[t]{0.45\linewidth}
        \centering
        \fbox{\includegraphics[width=\linewidth]{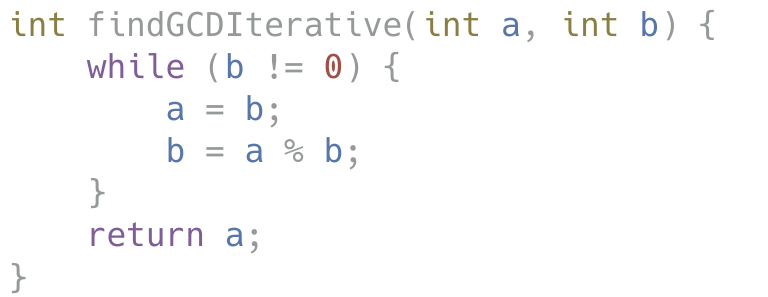}}
        \centerline{(b) Iterative Implementation}
    \end{minipage}

    \caption{Two Implementations of GCD}
    \label{fig:GCD_example}
\end{figure}

\noindent\textbf{Multiple Sequence Alignment Mechanism.}
To enhance the semantic features of token sequences, AlphaCC introduces the innovative concept of ``Code MSA’’, which retrieves lexically similar code fragments from a large codebase to align with the original token sequence. This approach significantly elevates the model's capacity to capture semantic information embedded in code.

Specifically, as illustrated in Figure~\ref{fig:GCD_example}, when presented with two code implementations for computing the greatest common divisor (GCD), one recursive and one iterative, the MSA mechanism can identify that these are semantic clones despite their structural differences. The critical operations in GCD functions revolve around the modulo operation \texttt{a\%b} and the base condition check \texttt{if(b==0)}. By incorporating similar code fragments that share these key operational characteristics but differ in other aspects, the MSA mechanism creates a rich learning environment where the model can distinguish between essential and non-essential code patterns. This data augmentation process, with its introduction of relevant variations, trains the model to focus on functionally significant code segments while ignoring irrelevant syntactic variations, thereby enhancing its sensitivity to semantically equivalent code despite structural dissimilarities.

\noindent\textbf{Dual-Attention Mechanism.}
The Codeformer employs a sophisticated dual-attention mechanism that operates at multiple levels of abstraction. The inner-sequence self-attention mechanism captures the relational dependencies between tokens within each individual token sequence, modeling control flow and data flow patterns. Concurrently, the inter-sequence cross-attention mechanism establishes connections across aligned positions in different sequences, enabling the model to identify patterns of token-level variations that preserve semantic equivalence.

This hierarchical attention framework allows AlphaCC to simultaneously reason about local code structures and their global functional implications~\cite{wang2016relation, zadeh2018multi}. For example, when processing different loop constructs that accomplish the same task, the inner-sequence attention captures the loop's internal logic, while the cross-attention recognizes the functional equivalence across different implementation styles. This multi-dimensional understanding of code significantly enhances the model's ability to detect semantically equivalent code fragments regardless of their syntactic representation.

\noindent\textbf{Streamlined Target of Token.}
AlphaCC leverages robust semantic modeling while maintaining superior computational efficiency and tool-independence compared to more complex approaches, such as IR-based methods~\cite{yu2019neural, zhang2019novel, xu2024dsfm, yu2023graph}. By directly analyzing code tokens, AlphaCC eliminates the need for third-party analyzers or intricate structural transformations. 
This streamlined approach offers strong adaptability, enabling effective application to diverse and complex scenarios.

\subsection{Threats To Validity}
\label{threats2validity}

Despite our rigorous evaluation of AlphaCC, several threats to validity arise due to practical limitations in resource accessibility.

First, our comparative analysis faces limitations in including certain baseline methods due to insufficient implementation details in original publications. For example, CC2Vec~\cite{dou2024cc2vec}, a token-based approach, could not be accurately replicated because the training methodology for token type weights is not explicitly described in the available literature. Similarly, DeepSim~\cite{zhao2018deepsim}, a graph-based method, lacks comprehensive details about its feature matrix transformation process, which hinders our ability to implement it faithfully. These information gaps prevent us from incorporating these potentially relevant methods as baselines in our experimental setup in a consistent way, which may affect the comprehensiveness of our comparative evaluation.

Second, our evaluation of AlphaCC relies predominantly on three widely recognized datasets: GCJ, BigCloneBench, and OJClone. Although these datasets represent standard benchmarks in code clone detection research, they may not fully capture the diverse spectrum of code instances encountered in real-world software development environments. The characteristics of these datasets might introduce biases that influence performance metrics. 
Furthermore, these benchmark datasets only contain limited programming language types, as additional datasets are not currently available. We hope that more benchmarks covering other languages will appear in the future, and we will extend our evaluation accordingly.


\section{CONCLUSION}\label{conclusion}

In this paper, we propose AlphaCC, a novel approach designed to tackle the challenges of limited capability for semantic code clone detection and strong reliance on third-party analyzers. 
AlphaCC represents each input code fragment as a token sequence. To enrich token semantic representation, AlphaCC constructs Code MSA to align lexically similar code fragments, thereby exposing their functional commonalities. The aligned token sequences are then processed by a dual-attention Codeformer module, which leverages inner-sequence self-attention to capture token-level dependencies within each individual token sequence, and inter-sequence cross-attention to model interactions across multiple aligned sequences, enabling a robust abstraction of underlying program logic. Finally, AlphaCC computes fine-grained token-level similarity between code fragments to effectively distinguish clone pairs. 
Comprehensive experiments on three diverse datasets demonstrate the tool-independence and superior semantic understanding of AlphaCC. Furthermore, AlphaCC achieves competitive efficiency, underscoring its practical utility for large-scale code clone detection tasks.

\section*{Data Availability}

The replication package, including code and datasets, is available at: \url{https://github.com/alphacc2025/alphacc}.

\bibliographystyle{ACM-Reference-Format}
\bibliography{ref}


\end{document}